\documentclass[fullpage,11pt]{amsart}
\usepackage[a4paper,left=2cm,top=2cm,right=2cm,bottom=2.5cm]{geometry}
\usepackage{amsfonts,amsmath,amsthm,array,amssymb}
\usepackage{hyperref}
\usepackage{cite}
\usepackage{indentfirst}
\usepackage{color}
\usepackage{fullpage}

\usepackage{xcolor}

\usepackage{graphicx,wrapfig}
\usepackage[font={small}]{caption}
\usepackage[font=footnotesize, labelfont=bf]{subcaption}

\title{Topological Methods for Characterising Spatial Networks: \\A Case Study in Tumour Vasculature}
\author{Helen M Byrne, Heather A Harrington, Ruth Muschel, Gesine Reinert, Bernadette J Stolz, Ulrike Tillmann}

\begin{document}  
\maketitle
\thispagestyle{empty}
\pagestyle{empty} 

\begin{abstract}
Understanding how the spatial structure of blood vessel networks relates to their function in healthy and abnormal biological tissues could improve diagnosis and treatment for diseases such as cancer. New imaging techniques can generate multiple, high-resolution images of the same tissue region, and show how vessel networks evolve during disease onset and treatment. Such experimental advances have created an exciting opportunity for discovering new links between vessel structure and disease through the development of mathematical tools that can analyse these rich datasets. 
Here we explain how topological data analysis (TDA) can be used to study vessel network structures. TDA is a growing field in the mathematical and computational sciences, that consists of algorithmic methods for identifying global and multi-scale structures in high-dimensional data sets that may be noisy and incomplete \cite{Edelsbrunner:2010va,Carlsson:2009gr,Zomorodian:2005wf}. TDA has identified the effect of ageing on vessel networks in the brain \cite{miller-brain-artery} and more recently proposed to study blood flow and stenosis \cite{nicponski2019topological}.
Here we present preliminary work 
which shows how TDA of spatial network structure can be used to characterise tumour vasculature.

\end{abstract}

\section{Introduction}

Angiogenesis, the formation of new blood vessels from existing ones, is the process by which a tumour establishes its own supply of nutrients so that it may grow and spread to other parts of the body. The mathematical community has been studying the mechanisms of tumour-induced angiogenesis for more than 35 years \cite{peirce2008,mantzaris2004,scianna2013}. In this time, there have been many advancements in obtaining high resolution, spatio-temporal data, improvements in computationally simulating data, while on the other hand, there has been the development of geometric/topological algorithms for analysing the shape of data. 

\subsubsection*{Tumour vessel networks.} Vessel network structure can reveal the presence of an underlying disease, or how a patient may respond to treatment. For example, tumour vasculature loses its hierarchical patterning, has different fractal dimensions, tortuous (\emph{i.e.} bendiness property) and enlarged vessels and different paths for blood flow, which can inhibit nutrient and drug delivery~\cite{Baish1996}. In addition, oxygen availability, which determines treatment responsiveness, depends on vessel network structure~\cite{Grimes2016}. Existing studies of vessel networks use metrics such as inter-vessel spacing, number of branching points, vessel length density, tortuosity and fractal dimensions~\cite{Tozer2005}. While these metrics can be related to tumour progression and treatment response~\cite{Jain1988}, their values may be sensitive to the algorithms used to convert the raw images to network based descriptions. Further, most measures characterise networks at a single spatial scale, although patterns may emerge at multiple spatial scales. TDA offers a promising and rigorous alternative for relating the structure of vessel networks obtained from raw images to their function and the disease status of the perfused tissue.

\subsubsection*{Mathematical models of artificial vessel networks.}
A large number of mathematical models have been developed over the past 30 years to study tumour-induced angiogenesis~\cite{peirce2008,mantzaris2004,scianna2013} and to simulate artificial vessel networks. For example, now we have multiscale, agent-based models that simulate angiogenesis and vascular tumour growth~\cite{Perfahl2016, Grogan2017a, Grogan2016b}. These models have been used to investigate how vessel networks respond to anti-angiogenic and other vascular targeting agents. For our purposes, we can simulate the models under different conditions, and obtain data, which we then aim to compare with experimental observations.


We remark that tumour-induced angiogenesis networks are inherently multi-scale ranging from arteries, to arterioles, to the smallest single-cell capillaries. Therefore it is natural to apply a multi-scale method for analysing such datasets. 

In this paper we present preliminary results where we apply TDA to tumor-induced vascular networks. First, we introduce topological data analysis and then describe the state-of-the-art angiogenesis data that are now being generated. Then we introduce angiogenesis data, and give preliminary results. This paper ends with a short discussion. 

\vskip .2in
\noindent
{\bf Centre for TDA:}
The authors are members of the Centre for Topological Data Analysis led by the second and last authors. The work here represents one of many lines of application driven research in the centre.

\section{Topological data analysis}

Data from biological processes which are observed as physical objects, such as the vascular networks of interest here, are inherently spatial networks. It is thus natural to employ data science methods to analyse the geometric features of the network.

\subsection{Mathematics}

One of the most developed tools in topological data analysis is persistent homology, our preferred method here~\cite{Edelsbrunner:2010va,Carlsson:2009gr,Zomorodian:2005wf}. 
We will describe below a generic procedure for how to associate to a data set its persistent homology barcode, its topological signature. As we will see, in applications this can flexibly be adjusted to suit the problem at hand. 

\subsubsection*
{From point clouds to topological spaces.} Given a point cloud $P$ in Euclidean space $\mathbb R^N$ 
we can define its $\epsilon$-neighborhood $N_\epsilon (P)$ to be the union of balls $B_\epsilon (p)$ of radius $\epsilon$  around all points $p$ in $P$: 
$$ N_\epsilon (P) = \bigcup_{p\in P} B_\epsilon ( p), \quad \quad \epsilon \geq 0.$$
The space $N_\epsilon (P)$ has topology that varies with $\epsilon \geq 0$  from a totally discrete space 
for $\epsilon =0$ to one large, featureless \lq blob' resembling a large ball for $\epsilon$ large.   
It is the spaces between these extremes that are of interest. We seek to determine how their characteristic topological features change as $\epsilon$ increases.

\subsubsection*
{From  topological spaces to combinatorial data.} Topological spaces can be approximated by combinatorial data, so-called simplicial complexes $K = \{ K_n\}_{n\geq 0}$. These are higher-dimensional analogues of graphs. Indeed, given a point cloud $P$ and $\epsilon \geq 0$, we construct the associated Vietoris-Rips complex $VR_\epsilon( P)$ by first  building a graph with vertices $P$ and edges $(p_0, p_1)$ for all pairs of points in $P$ of distance $d(p_0,p_1) \leq \epsilon$.
We then add an $n$-simplex for each complete subgraph on $n+1$ vertices.
The combinatorial data 
is thus  given by
$$
VR_\epsilon (P) = \bigcup _{n\geq 0} VR_\epsilon (P) _n, \quad \quad VR_\epsilon (P) _n = \{ (p_0, \dots , p_n)  \, | \,\, d(p_i, p_j) \leq \epsilon \text{ for all } i, j \}, $$
where $VR_\epsilon (P)_n$ lists all the $n$-simplices of 
our simplicial complex.

\subsubsection* 
{From combinatorial data to linear algebra.}

Consider a simplicial complex $K$, and let  $\mathbb F$ be a field; this could be the real or complex numbers but often for computational topology the field with two elements $\mathbb F_2 = \{ 0, 1 \}$ given by mod $2$ arithmetic is chosen.  Define the $n$-chains of $K$ to be the vector space
$\mathbb F [K_n]$ with basis the $n$-simplices, and a linear map $\partial_n: \mathbb F [K_n] \to \mathbb F [K_{n-1}]$ that takes  an $n$-simplex to its boundary, the (alternating) sum of its faces. For the Vietoris-Rips complex this is simply
$$
\partial_n (p_0, \dots , p_n) = \sum _{n\geq i \geq 0}
(-1)^i \, (p_0, \dots, \hat p_i, \dots , p_n) \quad \quad \text{where $\hat p_i$ denotes that $p_i$ is removed.}
$$
A straightforward computation shows that applying the boundary operator twice gives the zero map: $\partial_{n} \circ \partial_{n+1} =0$. This important algebraic identity reflects the geometric fact  that 
the boundary of a boundary is always empty.
Homology measures the difference between the cycles ($n$-chains with zero boundary) and the boundaries ($n$-chains that are boundaries of $(n+1)$-chains): For each $n\geq 0$ define the $n$-th homology group
$$
H_n (K) := \frac { \ker \partial_n  } {\text{im } \partial _{n+1} } \quad \quad b_n := \dim H_n(K)= \dim \ker \partial_n - \dim \text {im } \partial _{n+1}.
$$
The number $b_n$ is also called the 
{ $n$-th Betti number}:
$b_0$  is the number of connected components of $K$ and, for $K$  a graph,  $b_1$ is the total  number of  simple circuits. For more general $K$ the dimension of the higher homology groups give a count of how many higher dimensional cavities there are: For example, the boundary of an $(n+1)$-simplex has a non-trivial homology group 
in dimension
$n$ with Betti number $b_n =1$.

\subsubsection*
{Filtering and functoriallity.}
An important property of homology is  functoriallity: A map between simplicial complexes induces a map between their homologies. Now, for each $\epsilon' < \epsilon$ we have an  inclusion of the corresponding Vietoris-Rips complexes $VR_{\epsilon '} (P) \subset VR_{\epsilon } (P) $ and, by functoriallity, an induced  map in homology.
(Note, it is important here that we deal with homology groups and not just Betti numbers!) 
Thus with growing $\epsilon$ we can track elements in the homology groups of the corresponding complexes $VR_\epsilon (P)$. A non-zero element that first appears 
(\lq\lq is born")
at $\epsilon_{birth}$ and is mapped to zero (\lq \lq dies") at $\epsilon _{death}$ is represented by the interval (\lq \lq a bar") $[\epsilon_{birth}, \epsilon_{death})$. In $H_1$ this will correspond to the formation of a new circuit in the underlying graph at $\epsilon_{birth}$ and filling of that circuit with 2-dimensional simplices at $\epsilon_{death}$.
The $n$-th persistent homology $PH_n(P)$ of $P$  is the system of homology groups  $H_n(VR_\epsilon (P))$ for all $\epsilon >0$ with the induced maps.

\medskip 
There are two important results concerning persistent homology. 

\begin{figure}[htb!]
  \centering\includegraphics[width=3cm]{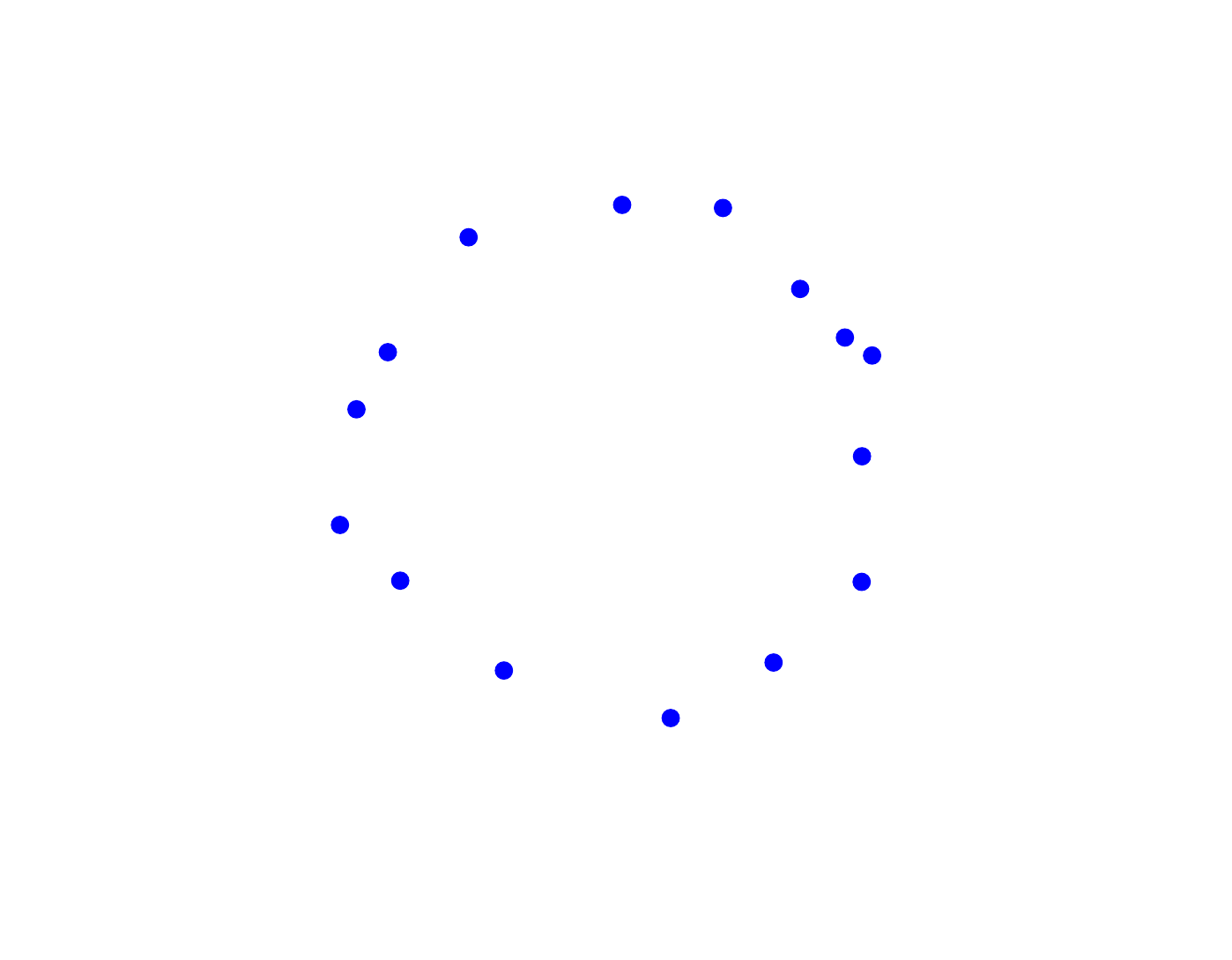}
\centering\includegraphics[width=3cm]{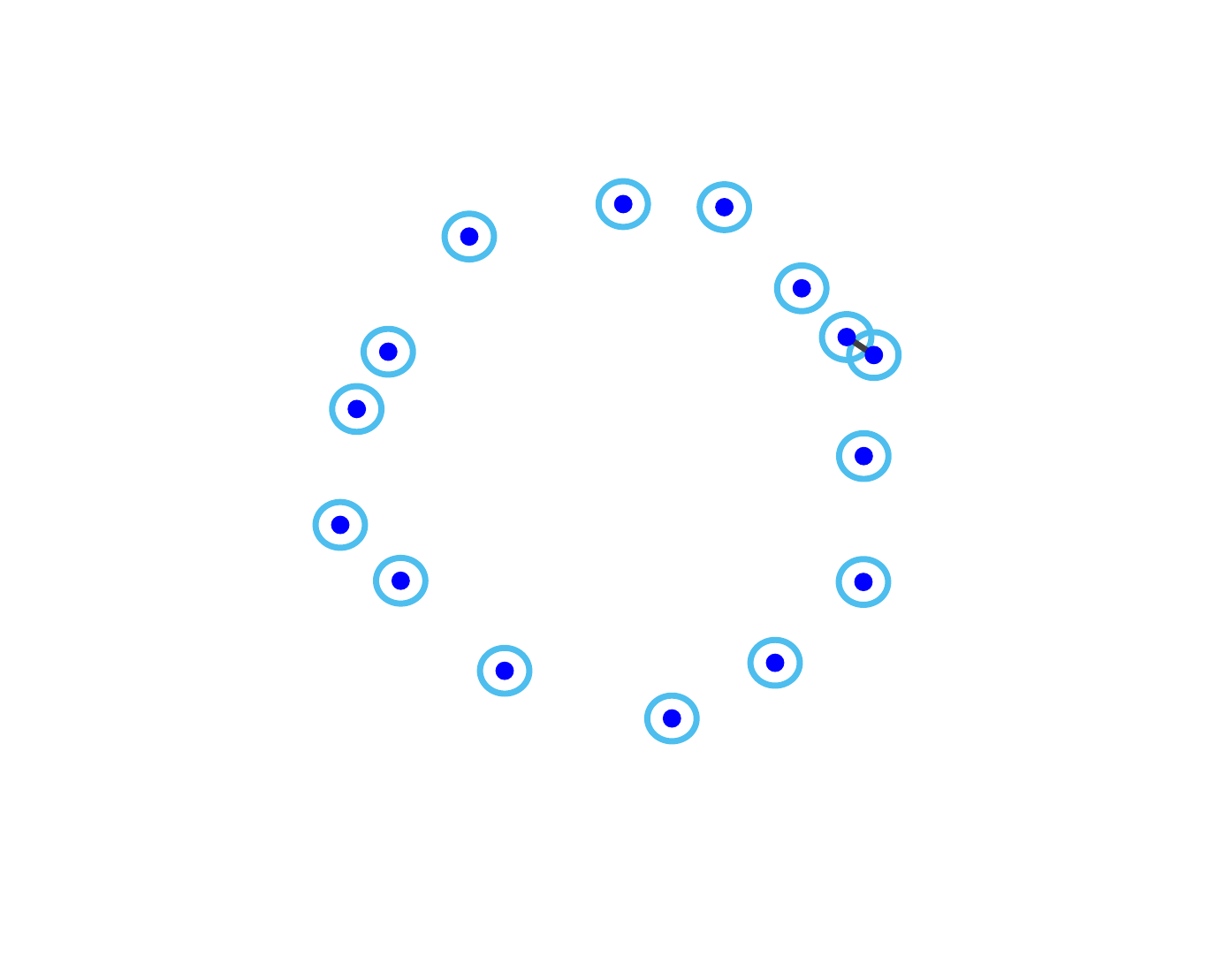}
\centering\includegraphics[width=3cm]{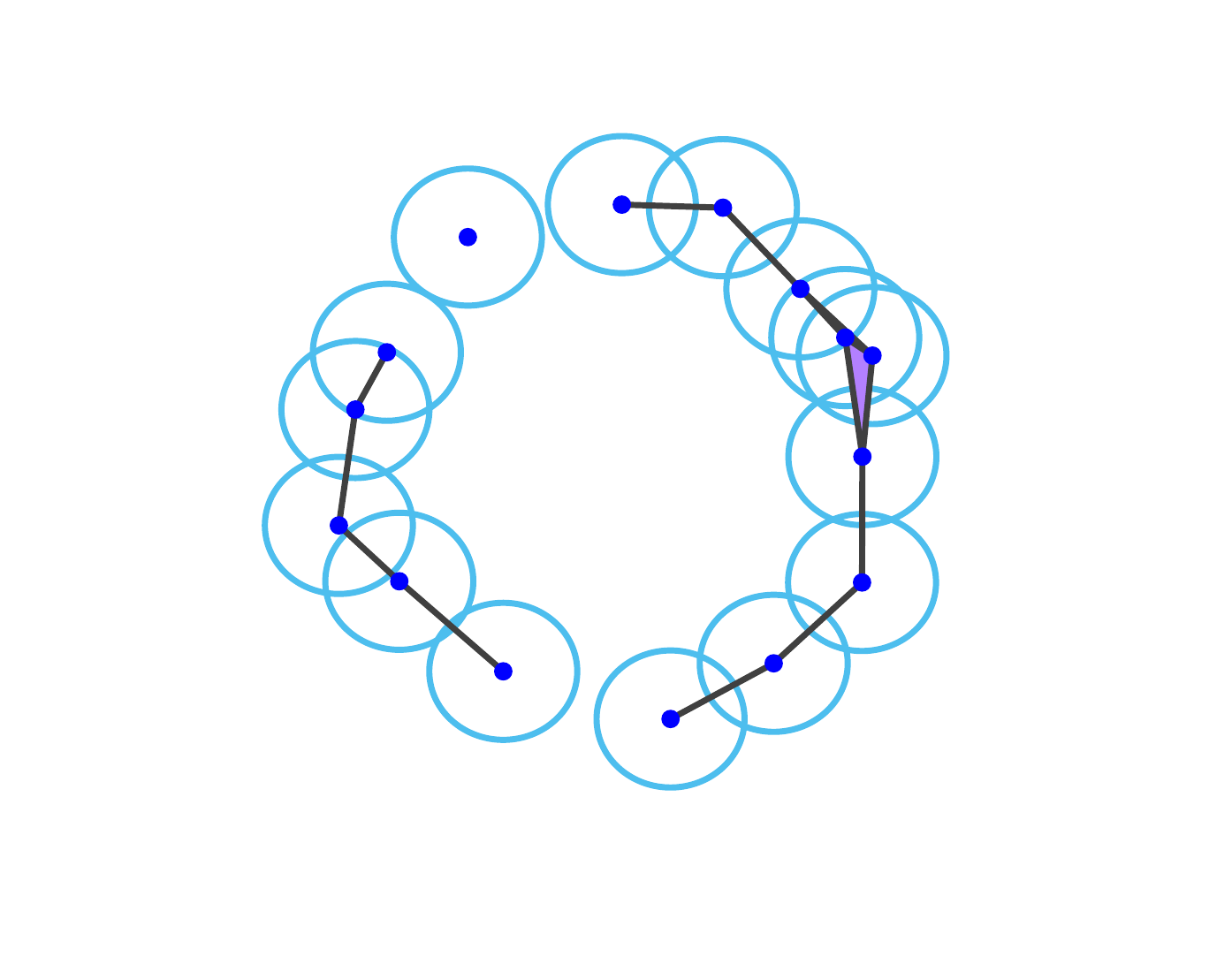}
\centering\includegraphics[width=3cm]{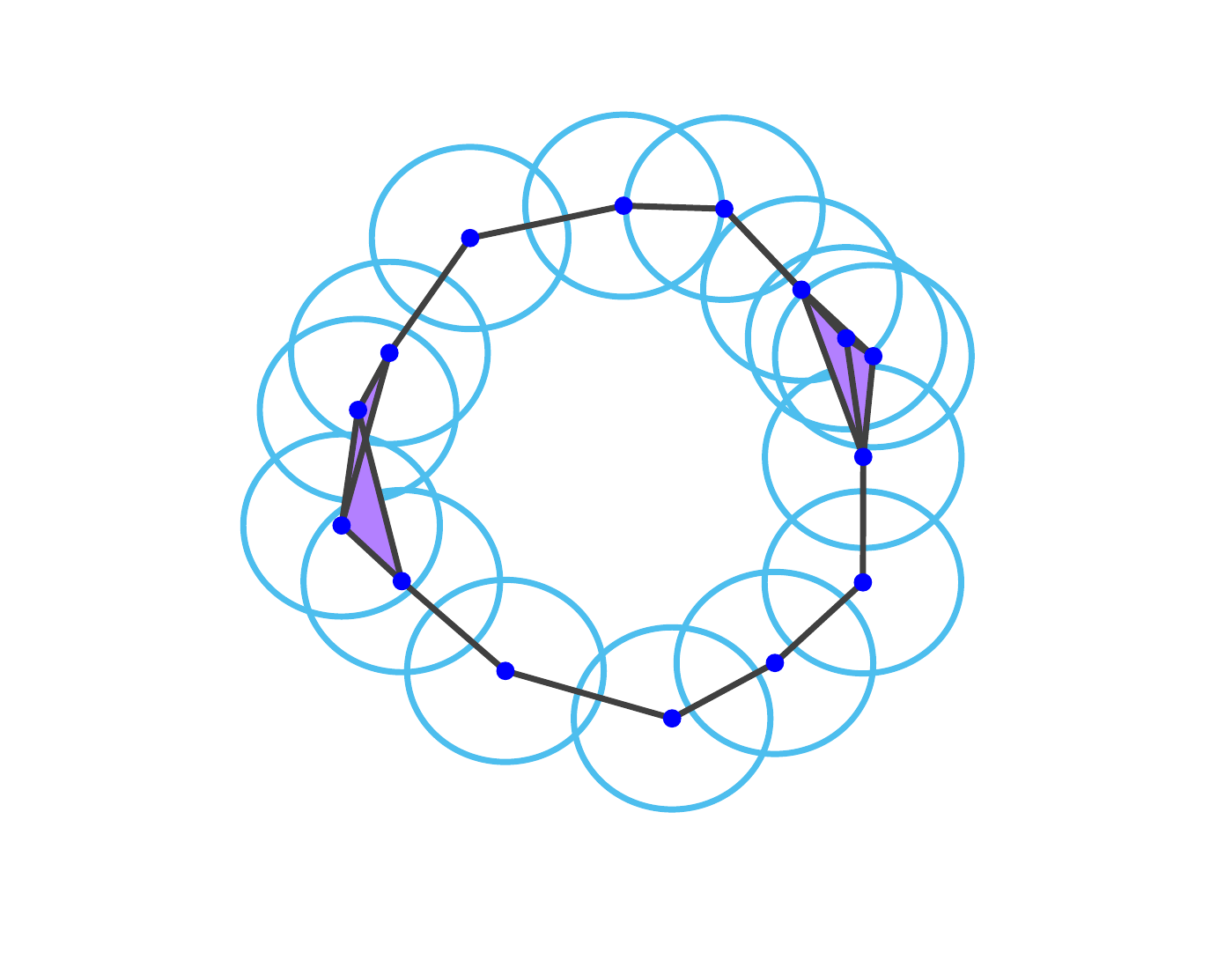}
\centering\includegraphics[width=3cm]{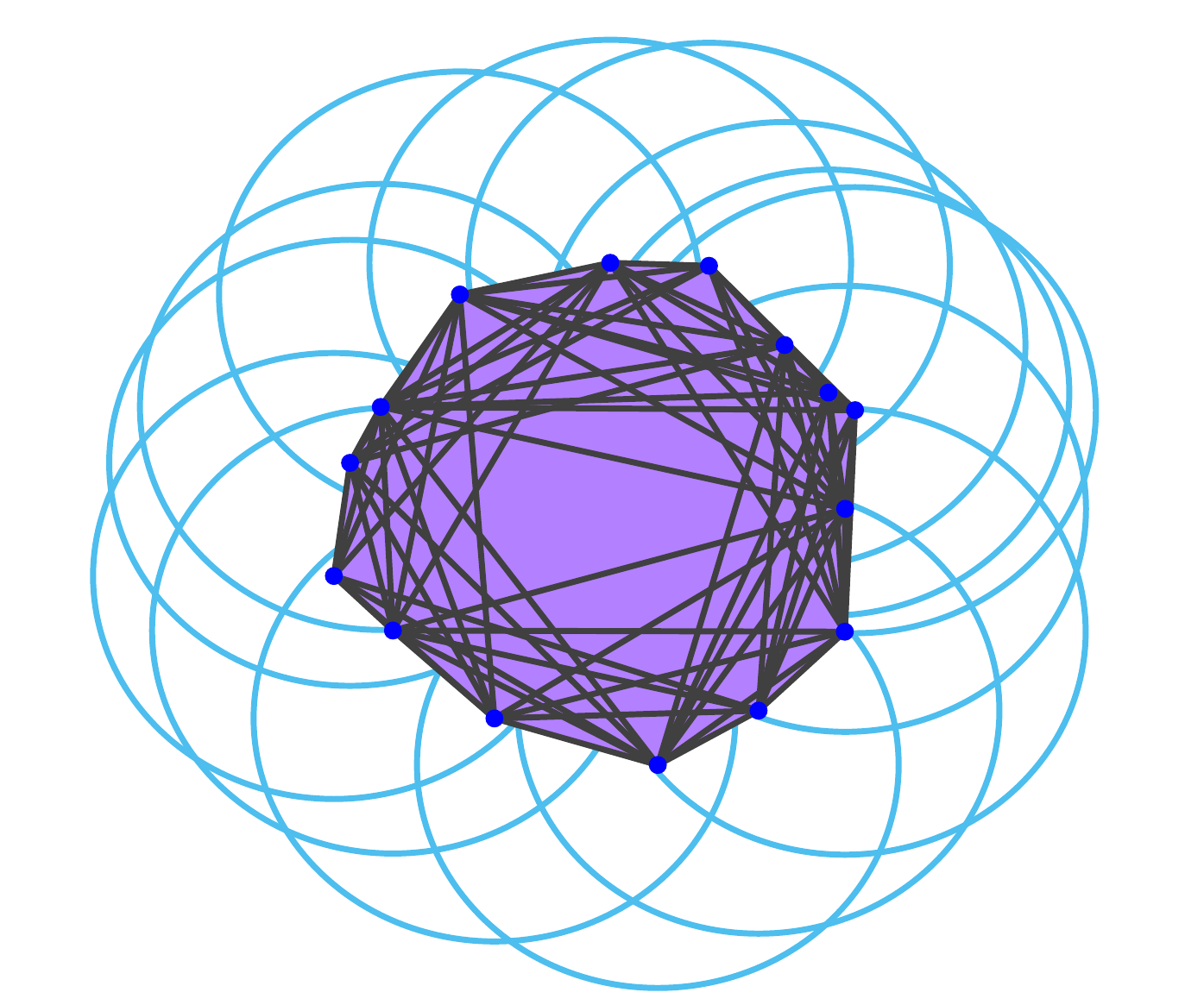}\\
\centering\includegraphics[width=.6\textwidth]{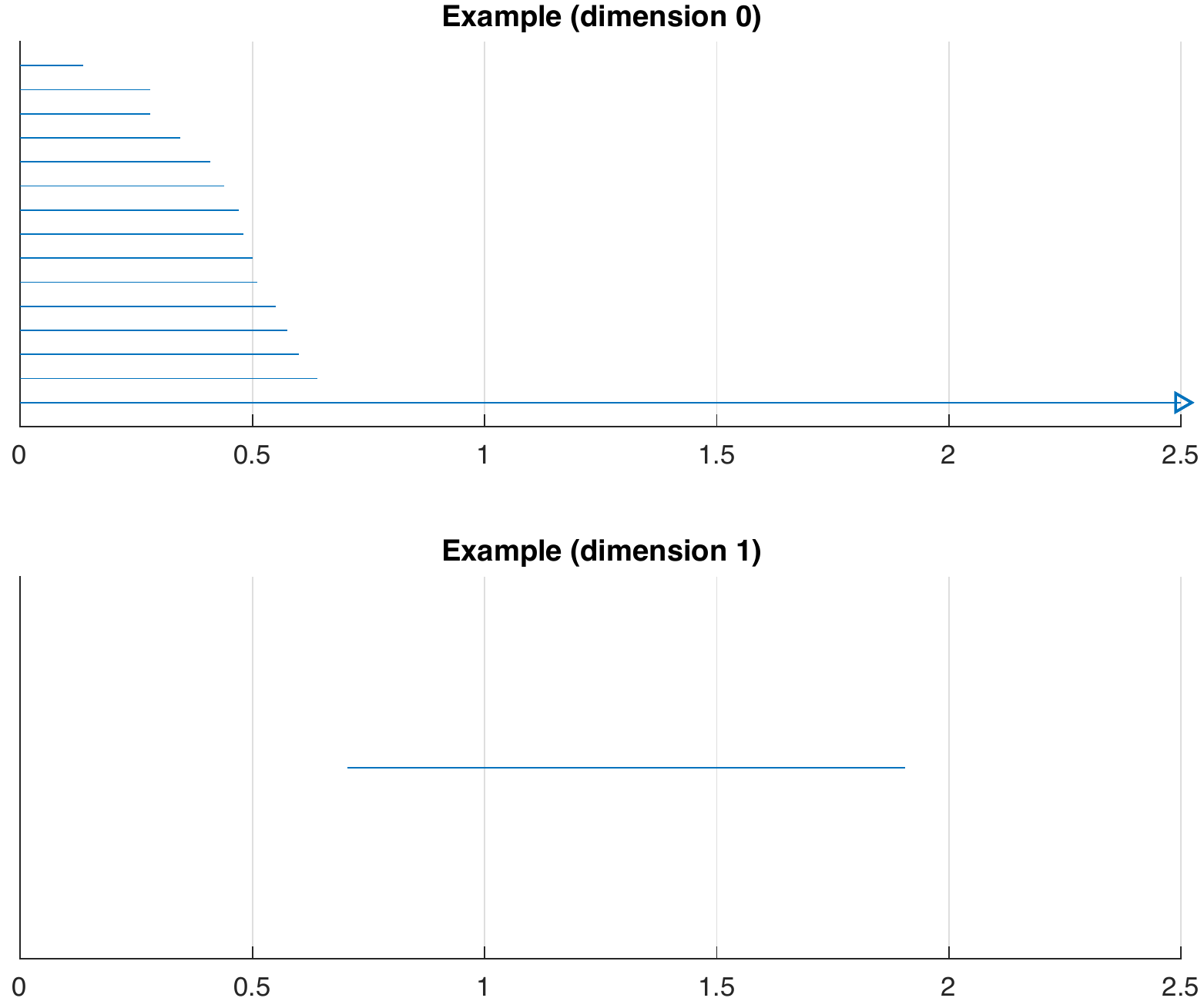} \caption{An example of a Vietoris-Rips filtration with examples of the complex at $\epsilon=0,0.1,0.55,1,2$ (top row). The corresponding barcodes in dimension 0 and dimension 1 are given where the horizontal axis gives the value of $\epsilon$.\label{fig:VR_Example}}
\end{figure}

\vskip .1in
\noindent
{\it Existence of Barcodes Theorem \cite{Carlsson:2009cq}.} For each homological dimension $n$,  compatible bases can be found so that for each $\epsilon$ the dimension of $H_n VR_\epsilon (P) $ is given by the number of corresponding bars containing $\epsilon$. In other words, the $n$-th persistent homology can be represented faithfully by a barcode; see Figure~\ref{fig:VR_Example}.

\vskip .1in
\noindent
{\it Stability Theorem \cite{Coehn-SteinerEdelsbrunnerHarer}.} If two  point clouds $P$ and $Q$ are close to each other (in the Hausdorff distance) then
the corresponding $n$-dimensional barcodes are also close (in the bottleneck distance). In other words, the persistent homology transform  $PH_n$ from point clouds to bars is continuous.

\vskip .1in
\noindent
{\it {Adapting persistent homology} to different data science problems.}
Not all data come in the form of a point cloud and different applications demand a change in the general set-up. The main feature of persistent homology is that it can track changing topology as the geometric object of interest is filtered in a suitable way. Above we considered a sequence of increasing Vietoris-Rips complexes as the radius of the little balls grew slowly until they eventually filled the whole space. Below we study larger and larger parts of the system of blood vessels in a tumour by growing the radius of vision (so just one ball) from the centre of the tumour.

\subsection{Statistical analysis in topological data analysis}
Once a Vietoris-Rips complex or a related TDA summary of the data, such as a barcode, is obtained, statistics can be used to detect deviations from what would be expected. Although for most summaries such as Betti numbers, theoretical results about what to expect at random are available, see for example \cite{kahle2014topology}, for most applications, these have too general assumptions and theoretical results of more restrictive models need yet to be derived.

In the absence of such results, simulations are employed for statistical inference. Suppose that data from a candidate 
model can be generated and their TDA summaries calculated. These summaries can then be used to indicate what is to be expected under the model. An observed summary outside the range, or close to the boundary of this range, of these simulated summaries, indicates deviation from the expectation. 
For vascular networks, theoretical models have been investigated for example in \cite{Perfahl2016}, but depending on the complexity of the simulations, simpler models based on branching processes may be more appropriate.
If it is not possible to simulate from appropriate models, then statistical machine learning offers a model-free approach to classify data through using TDA summaries as features in an automated learning method such as random forests.

\subsection{Computation.}
Over the last decade, many efforts have been made to develop and improve the computation of persistent homology.
The first paper on the standard persistent homology algorithm was introduced in \cite{zomorodian2005}. Since then, multiple software libraries have been developed that enable different types of data to be analysed, for example, point cloud, networks or images, as well as analyse data with different filtrations.
The data structure input and output differ for each library. There is also software for computing statistical summaries of topological outputs within many of the libraries. For a review of software and tutorial, see \cite{otter2017}.

\section{Angiogenesis data}

\begin{figure}
\subcaptionbox{}{\centering\includegraphics[height=6.3cm]{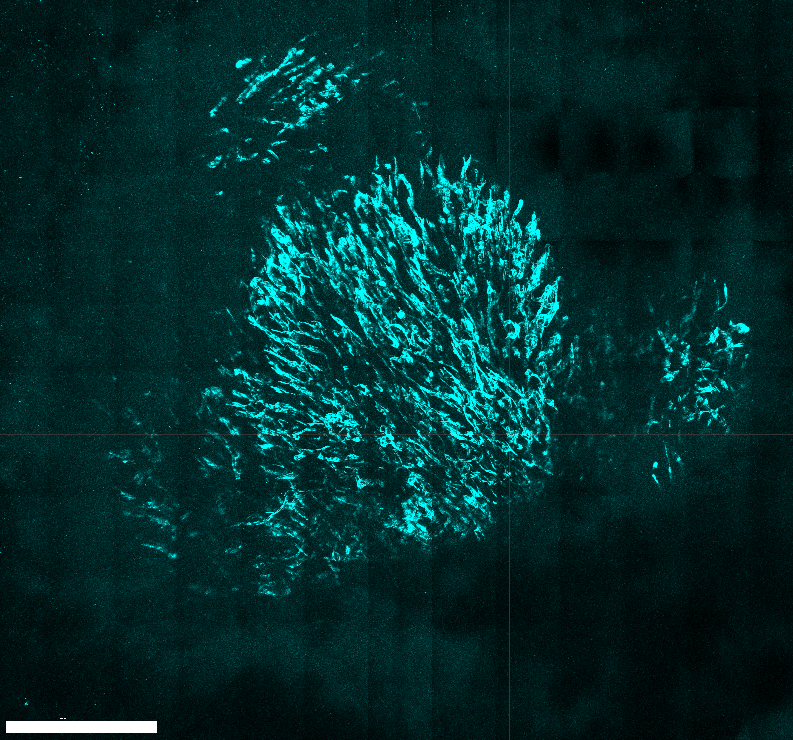}}%
\hspace{0.02\textwidth}
\subcaptionbox{}{\centering\includegraphics[height=6.3cm]{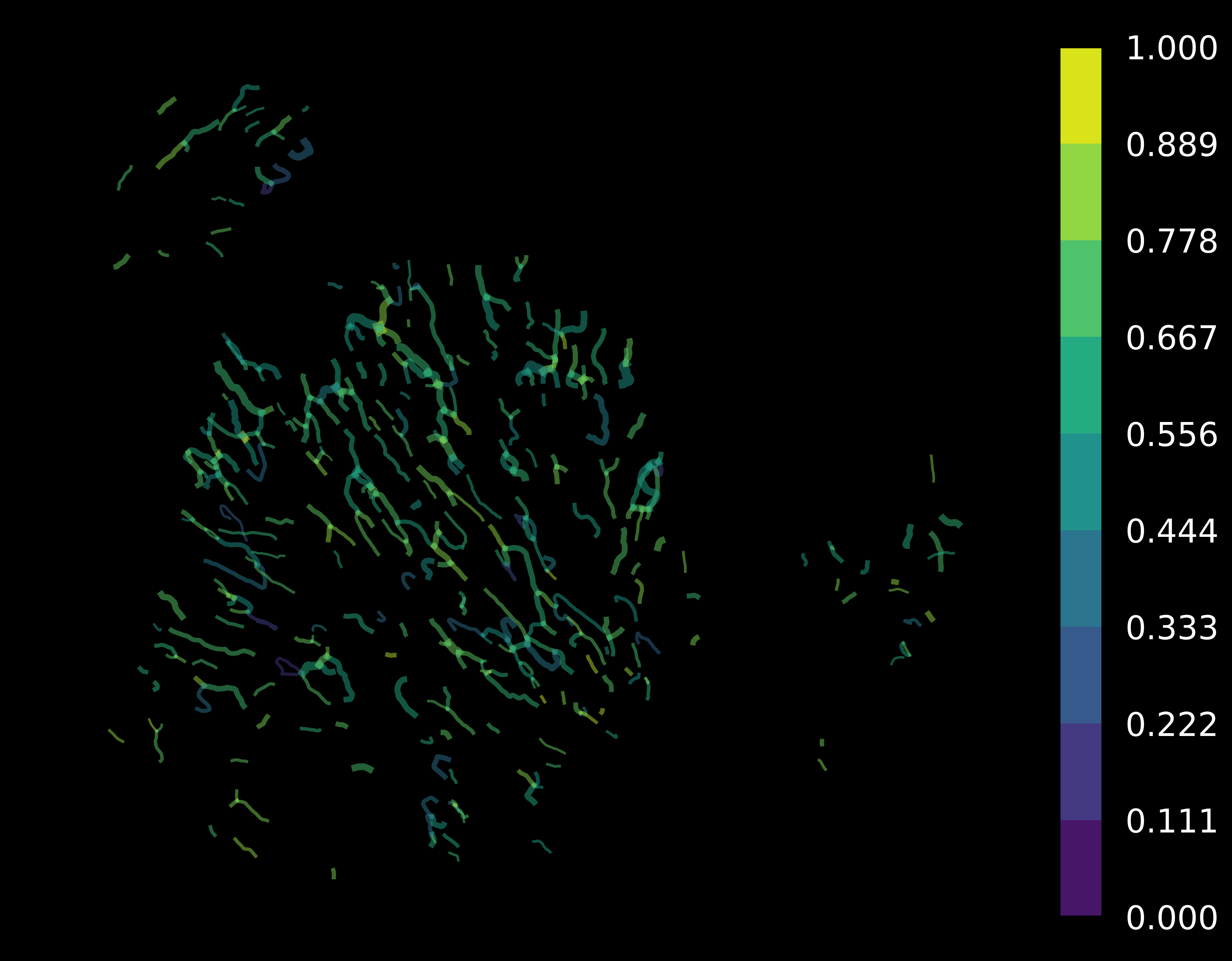} }%
\\
\subcaptionbox{
}{\centering\includegraphics[width=0.45\textwidth]{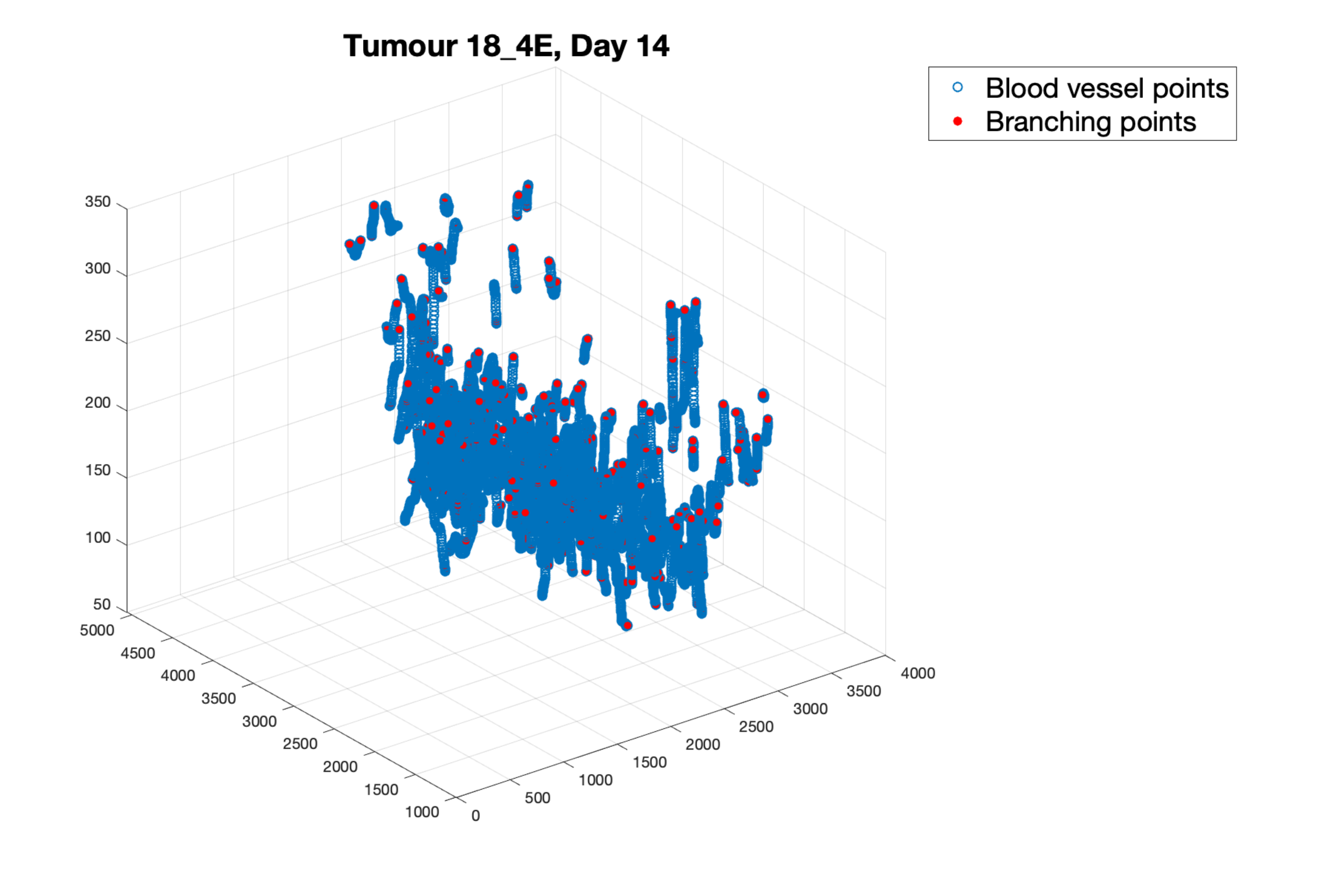}}%
\hspace{0.02\textwidth}
\subcaptionbox{}{\centering\includegraphics[width=0.48\textwidth]{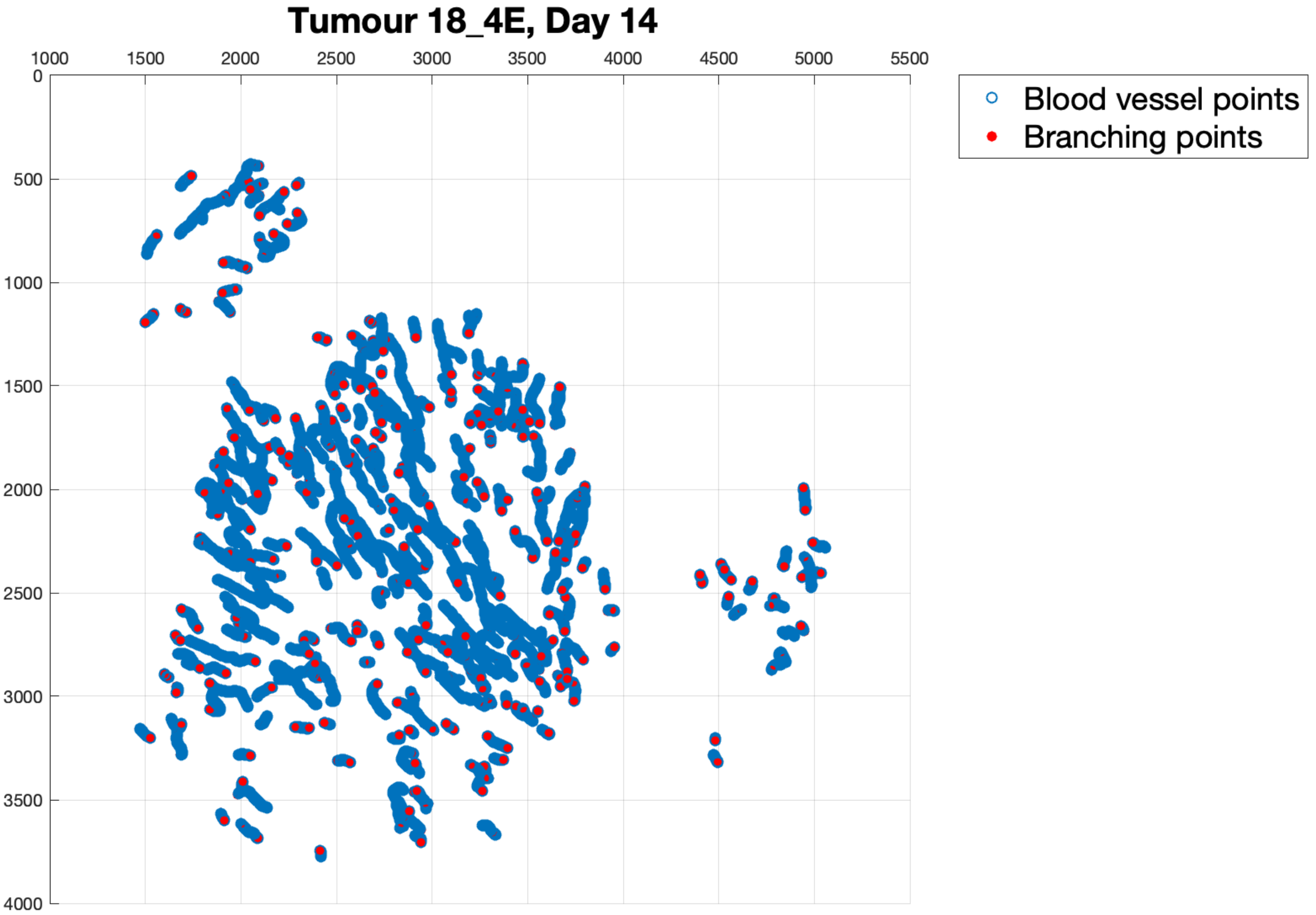} }%
\caption{Example images of tumour blood vessels. (a) Tumour blood vessels as seen under the microscope. The view corresponds to the first 2D slice of the tumour in the $z$-direction, the grey bar indicates the length of 1000$\mu$m in the $xy$-plane. The image was taken by Bostjan Markelc and Jakob Kaeppler. (b) 2D perspective on the extracted 3D skeleton of tumour blood vessels coloured according to a measure of tortuosity (chord-length-ratio). The 3D skeleton was created using all 2D microscopy image slices that were obtained from the tumour. The skeleton and tortuosity values were extracted by Russel Bates. (c) Vessel points we extracted from the 3D skeleton. 
We show branching points, i.e. vessel points at the beginning or end of a branch, in red. We interpret all blood vessel points, i.e. both branching points and non-branching points, as nodes and physical connections between them as edges in a blood vessel network. Blood vessel networks are the input into our topological analysis.
(d) Perspective corresponding to the 2D image slice shown in (a) and (b) on vessel points we extracted from the skeleton image. 
}\label{fig:ExampleImages}
\end{figure}

 High-resolution 3D images showing vessel network evolution in mouse tumours undergoing treatment (with chemo-, radio- and anti-angiogenic therapy) are being generated by Prof. Ruth Muschel’s research group (see Fig.~\ref{fig:ExampleImages}(a) for an example), while Prof. Mike Brady’s group (Biomedical Engineering) are developing new image segmentation tools to extract vessel structures from these images [6] see Fig.~\ref{fig:ExampleImages}(b) for an example of extracted structures). The imaging data here were obtained from tumours generated from murine colon carcinoma cells in mice genetically engineered to have fluorescent endothelium. Using video 2-photon microscopy the entire tumour and its blood vessels were visualized daily after tumour induction. The mice were subjected to different treatment regimens once the tumours reached a specified size: (1) Controls. (2) Treatments known to increase vessel sprouting. (3) Treatments known to decrease vessel sprouting. (4) Treatment by single-dose irradiation (1 $\times$ 15 Gy). (5) Treatment by fractionated-dose irradiation (5 $\times$ 3 Gy). In addition to examining the structure of the vessels, vascular function was also evaluated.  While statistical techniques can characterise facets of the networks (\emph{e.g.} vessel lengths, radii and tortuosity), analysis and interpretation of their topological features and how these vary across spatial scales remain open problems which TDA is ideally suited to address.

\section{Preliminary results}

We characterise the unique features of tumour blood vessels, in particular the loops and the high degree of tortuosity, using persistent homology. We exploit the fact that our data are already in the form of a network rather than a point cloud.

Loops can be captured by persistent homology in dimension 1 of any filtration that is built using the structure of the vessel network data. Tortuosity has previously been successfully quantified studying persistent homology in dimension 0 of a filtration that can be imagined as a step-wise sliding of a plane over a biological network: In the first filtration step the entire network is situated on one side of the plane. As the plane moves, it starts intersecting the network until eventually the whole network is located on the other side of the plane. The side of the plane that is initially empty thereby gives rise to a sequence of embedded objects that can be interpreted as a filtration of the network. 
This approach has been used to quantify 
the tortuosity of brain arteries~\cite{Bendich2014} and the geometric structure of airways~\cite{Belchi2018}. Kanari \emph{et al.}\cite{Kanari2018} use a similar approach to classify the branching patterns of neurons using radial distances from the neuronal tree root, \emph{i.e.} considering a sphere with decreasing radius around the root.   These examples motivate the approach adopted here.

\begin{figure}[ht!]
\centering
\includegraphics[width=\textwidth]{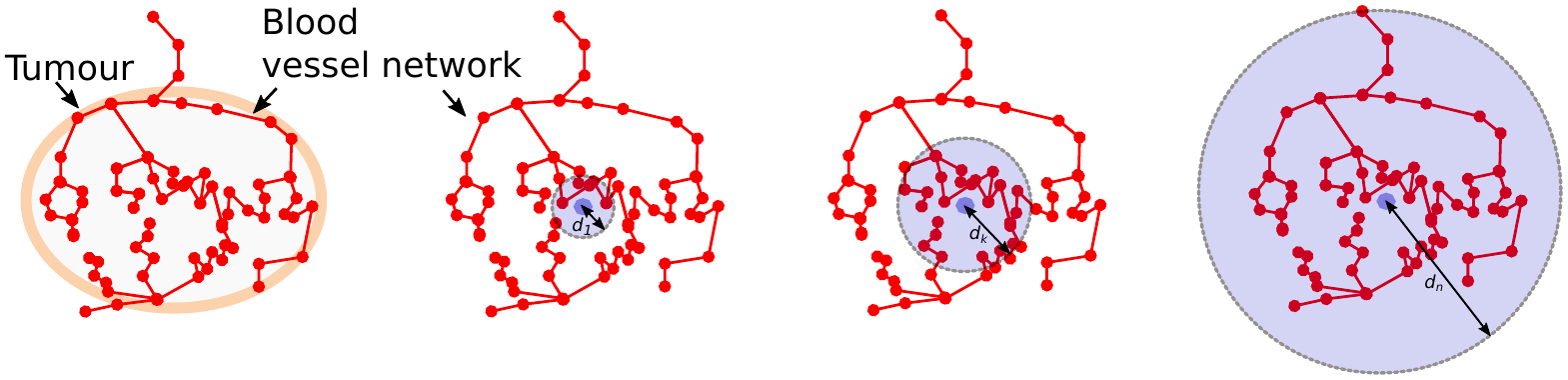}
\caption{Schematic illustration of the radial filtration of a tumour blood vessel network. On the $k$-th filtration step we include all vessel nodes and edges that are fully contained in the purple ball of radius $d_k$ around the centre of mass of the vessel points.}\label{RadialFiltrationSteps}
\end{figure}

In contrast to brain arteries or neurons, tumour blood vessels are not tree-like objects and they do not have a natural orientation.
By exploiting the fact that tumours are often viewed as spherical objects, we root our filtration in the tumour centre. Since we perform our analysis on the tumour blood vessels rather than the tumour itself, we approximate the tumour centre by the centre of mass of the blood vessel point data.
We then search the neighbourhood of the centre point, increasing the radial distance step-wise to include all vessel points within the radius. If two points that are connected by an edge in the blood vessel data are within the given radius, we add the edge to our filtration. Fig.~\ref{RadialFiltrationSteps} shows a schematic of the radial filtration on blood vessel data. Based on this filtration we study the topology of the growing network at every filtration step capturing tortuosity in dimension 0 and loops in dimension 1. Since the filtration generates topological information with respect to the tumour centre, we also obtain information about the heterogeneity of these characteristics within a single network. 

\begin{figure}[ht!]
\centering
\subcaptionbox{}{\includegraphics[width=0.49\textwidth]{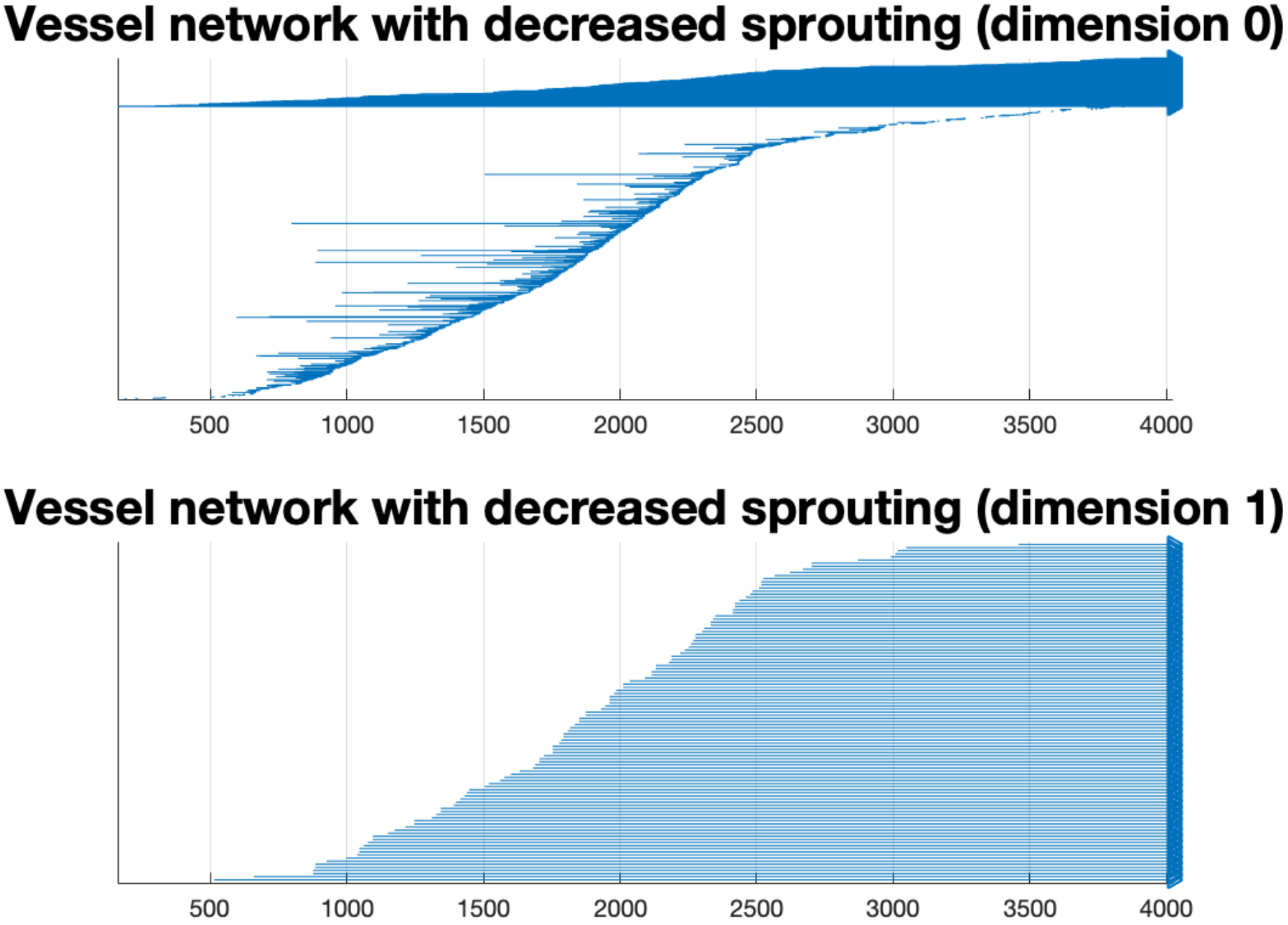}}%
\hspace{0.01\textwidth}
\subcaptionbox{}{\includegraphics[width=0.49\textwidth]{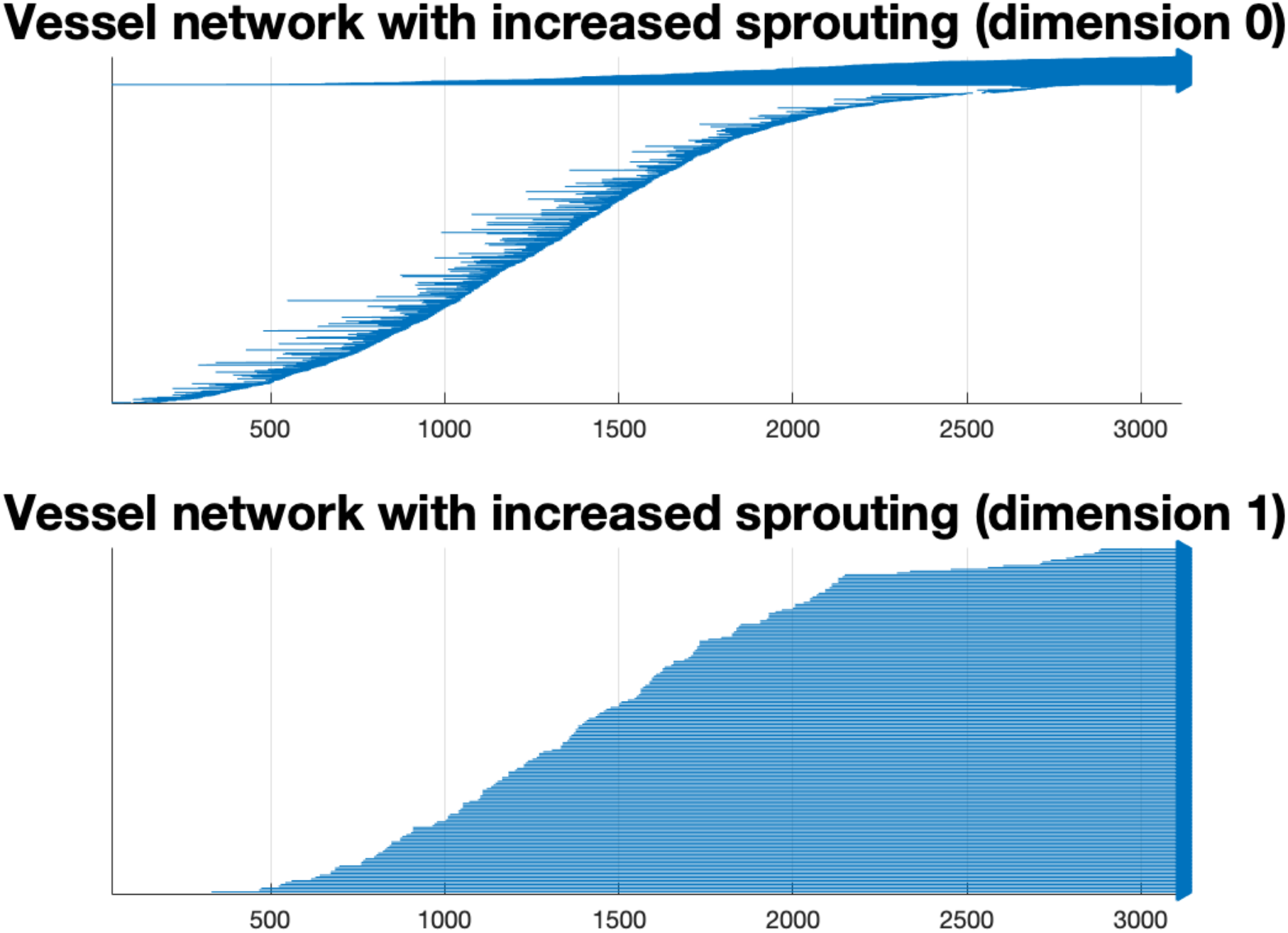}}%
\caption{Example barcodes from the radial filtration of networks subjected to two different treatment conditions. The vessel networks were imaged three days after treatment was administered. The top row shows features in dimension 0, i.e. connected components, the bottom row shows loops in the networks. Every bar in the barcodes represents one topological feature as the radius increases (horizontal axis). The beginning of the bar corresponds to the distance to the tumour centre at which the feature is first recorded in the filtration, the end of the bar corresponds to the distance to the tumour centre when the feature disappears in the filtration.}\label{Fig:ExampleBarcodes}
\end{figure}

In Fig.~\ref{Fig:ExampleBarcodes} we present barcodes obtained from the radial filtration of tumour blood vessel networks that were subjected to different treatment regimes: (a) treatment to decrease vessel sprouting and (b) treatment to increase vessel sprouting.

We observe differences in the barcodes in both dimensions. We note also that the barcodes end at different filtration points, \emph{i.e.} around 4000 for the decreased sprouting case and around 3000 for the increased sprouting case. These values reflect biological differences in the sizes of the two tumours. In dimension 0, where we expect to capture tortuosity, we find short-lived connected components, \emph{i.e.} short bars, interspersed with longer-lived, connected components, emph{i.e.} more persistent bars. In dimension 1, we see more loops in the network exposed to treatment that increases sprouting than the network exposed to treatment that decreases sprouting, as expected. This indicates that barcodes can capture the effects of treatment on both the sprouting behaviour and tortuosity of vessel networks.

\section{Discussion and outlook}

In this article we have presented preliminary results which show how topological data analysis (TDA) can be used to quantify changes in the morphology of tumour vascular networks following exposure to treatments which alter the rate at which new vessels form. By employing TDA, we could analyse the vessels at multiple scales from the tumour centre, evading selecting a threshold for network analysis. Thus, TDA also can offer a method for detecting parameter sensitivity in topological features, which may be useful for other multiscale datasets. While at this stage, differences in the spatial distribution of tortuosity or loops are not apparent, in future work, we will investigate these observations further by applying statistical analyses to the full dataset. 

At the same time, we will use our existing multiscale models of vascular tumour growth to generate artificial networks for analysis with TDA~\cite{Perfahl2016, Grogan2017a, Grogan2016b}. 
In this way, we aim to establish whether observed patterns (such as persistence of cycles in the topological description of the network) can be related to specific biophysical mechanisms used to construct the vessel networks, to investigate how the barcode associated with a particular tumour changes as the tumour and its vasculature evolve.
We will also simulate the experimental image acquisition process by adding noise associated with multi-photon microscopy to the artificial networks. Analysis of the synthetic networks will be used to assess the robustness of TDA to image-processing artefacts.

In the longer term the application of TDA to clinical images and spatial biomedical networks may reveal new relationships between network structure and treatment response. For example, in cancer, there is no consensus about 'vessel normalization theory'~\cite{Jain2005}, which proposes that vascular targeting agents transiently improve tumour response to therapy by reverting an aberrant tumour vasculature to one resembling healthy tissue. TDA may be used to determine if characteristics more typical of healthy vasculature are being restored during treatment, thereby improving the diagnostic potential of existing imaging methods. In the future, we will also investigate TDA methods for other applications in wound healing, retinal pathology and cardiac disease.

\section{Acknowledgements}

We would like to thank Russel Bates, James Grogan, Bostjan Markelc, Jakob Kaeppler and Nicola Richmond for helpful discussions.
BJS gratefully acknowledges the EPSRC and MRC (Grant number EP/G037280/1) and F. Hoffmann-La Roche AG for funding her doctoral studies. HAH gratefully acknowledges funding from a Royal Society University Research Fellowship. The Centre for Topological Data Analysis is funded by the EPSRC
(Grant number EP/R018472/1). 


\bibliographystyle{siam}
\bibliography{Vessels}

\end{document}